# Revealing pre-earthquake signatures in atmosphere and ionosphere associated with 2015 M7.8 and M7.3 events in Nepal. Preliminary results

Dimitar Ouzounov[1], Sergey Pulinets[2], Dmitry Davidenko[2]

[1]Center of Excellence for Earth Systems Science & Observations, Chapman University, CA, USA
[2]Space Research Institute, Russian Academy of Sciences, Moscow, Russia

Correspondence to: Dimitar Ouzounov (dim.ouzounov@gmail.com)

**Abstract**

We analyze retrospectively/prospectively the transient variations of three different physical parameters of atmosphere during the time of M7.8 and M7.3 events in Nepal: outgoing earth radiation (OLR), GPS/TEC and the thermodynamic proprieties in the lower atmosphere. We found that in mid March 2015 a rapid augment of satellite observed earth radiation in atmosphere and the anomaly located in close vicinity to the future M7.8 epicenter reached the maximum on April 21-22. Our continuous satellite analysis revealed prospectively the new strong anomaly on May 3th, which was the reason to contemplate another large event in the area. On May 12, 2015 a large aftershock of M7.3 occurred. The analysis of air temperature from weather ground station near Katmandu shows analogous patterns with offset 1-2 days earlier to the satellite anomalies. The GPS/TEC data analysis indicates an augment and variation in electron density reaching a maximum value during April 22-24 period. A strong negative TEC anomaly in the crest of EIA (Equatorial Ionospheric Anomaly) has occurred on April 21[st] and strong positive on April 24[th], 2015. Our preliminary results show correlation between the pre-earthquake atmospheric and ionospheric anomalies and the occurrence of 2015 M7.8 and M7.3 events in Nepal.

## 1. Introduction

The observational evidence, from the last three decades studies in the different parts of the world provide a significant pattern of transient anomalies preceding earthquakes [Tronin et al., 2002; Liu et al., 2004; Pulinets and Boyarchuk, 2004; Ouzounov et al, 2007; Němec et al., 2008, 2008; Parrot, 2009; Kon et al., 2010; Hayakawa et al., 2013; Tramutoli et al., 2013]. Despite these results, there is still lack of consistent data observed prospectively that could help to revile the connection between atmospheric and ionospheric variations (or anomalies) associated with major earthquakes.

The recent M7.8 Nepal earthquake of April 25, 2015 was the largest recorded earthquake to hit this nation since 1934 (Figure 1). In this work we analyzed ground and satellite data to study the atmospheric and ionospheric response to the latest M7.8 and M7.3 in Nepal in 2015. We provided analogous study after Tohoku earthquake in Japan, which was done fully retrospectively. The difference this time is that immediately after M7.8 of April 25, in Nepal, we could analyze and find anomalies in atmosphere prospectively and to acknowledge in advance the potential for occurrence of M7.3 of May 12, 2015 [Ouzounov and Pulinets, 2015].

We examined three different physical parameters characterizing the state of the atmosphere/ionosphere during the periods before and after the event: 1. Outgoing Long wave Radiation, OLR (infra-red 10-13 μm) measured at the top of the atmosphere; 2. GPS/TEC (Total Electron Content) ionospheric variability; and 3. Air Temperature and Relative humidity from NOAA ground stations.

## 2. Observations and Analysis

### 2.1 Earth radiation observation

One of the main parameters we used to characterize the earth's radiation environment is the outgoing long-wave-earth radiation (OLR). OLR has been associated with the top of the atmosphere integrating the emissions from the ground, lower atmosphere and clouds [Ohring G. and Gruber, 1982] and primary was used to study Earth radiative budget and climate [Gruber and Krueger, 1984; Mehta and Susskind, 1999].

A daily OLR data were used to study the OLR variability in the zone of earthquake activity [Liu, 2000; Ouzounov et al., 2007, 2011; Xiong at al., 2010]. An augment in radiation and a transient change in OLR was proposed to be related to thermodynamic processes in the atmosphere over seismically active regions. An anomalous character of this was defined by [Ouzounov et al., 2007] as statically defined maximum change in the rate of OLR for a specific spatial location and predefined times and was constructed analogously to the anomalous thermal field proposed by [Tramutoli et al., 1999, 2013].

Our preliminary results show that in mid March 2015 a rapid augment of transient infrared radiation was observed from the satellite data and an anomaly near the epicenter reached the maximum on April 21-22, three days before the M7.8 (Figure2). The ongoing prospective analysis of satellite radiation revealed another transient anomaly on May 3-4 (8 days in advance), associated with the M7.3 of May 12, 2015 earthquake.

The time series plot for anomalous OLR (Figure 3,top, red), within 200 km radius around M7.8 epicenter for the period of Jan 1 – May 31, 2015 had confirmed that the maximum change in the OLR state over the epicentral area did occur in two periods, during April 21-22 and May 3-4. For comparison, the same analysis for 2014 (year, with no major seismicity) show (on Figure 3, top, black) low intensity of OLR level and absence of elevated values in comparison to the same of 2015.

This rapid enhancement of radiation could be explained by an anomalous flux of the latent heat over the area of increased tectonic activity. Analogous observations were observed within a few days anterior to the most recent major earthquakes Japan (M9, 2011), China (M7.9, 2008), Italy (M6.3, 2009), Samoa (M7, 2009), Haiti (M7, 2010) and Chile (M8.8, 2010) [Ouzounov et al., 2011a,b, c; Pulinets et al., 2015].

**2.2 Air Temperature Observation**

Multi year day-by-day counts of nighttime temperatures were used to compute the daily temperature variations near the vicinity of earthquake epicenter. Air temperatures data near the ground surface were obtained from Tribhuvan International Airport, Nepal thru NOAA Surface Data Hourly Global Database. We analyzed surface air temperature nighttime data for the period of 2011-2015 close to the terminator time 0500-0600 LT to define the normal and abnormal state of the air temperature. The time series for January 1-May 31, 2015 is shown in Fig.3B. We computed the residual values as a distinction between current value and the multiyear year mean of air temperature variability. The maximum offsets from the mean value reached near +5C on April 20 and +4C on May 2 and April $5^{th}$ (Fig.3B) with confidence level of more than +2 sigma for the all the observations from 2011 to 2015. The transient rapid augment of the surface air temperature is little ahead then the satellite remote observations showed on Fig 3A which is in agreements with the thermodynamic processes explained by LAIC concept. [Pulinets and Ouzounov, 2011].

**2.2 Ionospheric observation**

Ionospheric effects detected over the earthquake preparation zone of the Nepal M7.8 earthquake are very analogous to those detected before the strong earthquakes in China (Wenchuan M7.9 of May12, 2008 and Lushan M7.0 of April, 20 2013) [Pulinets et al., 2010]. For our analysis of ionospheric data we used two sources of information: global maps of the total electron content as IONEX index provided by JPL, and time series of calculated vertical TEC of lhaz GPS/GLONASS receiver in the region.

We computed GPS/TEC differential maps with exclusion day of 101 (April 11) what we considered is a magnetic storm day. The most unusual are the days 111 and 114 (Apr 21 and 24), when we may contemplate the negative and positive anomalies. To check this supposition the differential GIM maps were built for these days, which are presented in the Figure 4.

It could be seen strong negative (in EIA crests) anomalies on 21 April and very strong positive anomaly on 24 April. The effect reflects the equatorial anomaly strength change. Considered that epicenter is inside the northern crest of EIA we will express the strength of the anomaly as relation of TEC at northern crest to the TEC in the trough of EIA.

From the historical data of ionosphere monitoring was elaborated conception of the precursor mask, which generalizes the pattern of ionosphere parameter variations (foF2 or GPS TEC) few days before the earthquakes [Pulinets et al., 2002, 2014; Davidenko, 2013]. We used this approach to analyze the ΔTEC variations for the lhaz GPS/GLONASS receiver for the period around the time of the two Nepal earthquakes what is presented in the Figure 5. We found an appearance of nighttime positive deviations before both strong earthquakes lasting more than one day and repeating every day at the same local time. The positive variations start at 6PM and finish at 6 AM and equal to ionospheric precursor mask [Davidenko, 2013; Pulinets et al., 2014], could be connected with the terminator time for Nepal [Zolotov, 2015].

## 3. Summary and Conclusions

The joint analysis of atmospheric and ionospheric parameters during the M7.8 earthquake in Nepal has demonstrated the presence of correlated variations of atmospheric anomalies implying their connection with the earthquake preparation processes. One of the possible explanations for this relationship is the Lithosphere- Atmosphere- Ionosphere Coupling mechanism [Pulinets and Boyarchuk, 2004; Pulinets and Ouzounov, 2011, Pulinets et al., 2015], which provides the physical links between the different geochemical, atmospheric and ionospheric variations and tectonic activity. Our results show evidence that processes related to the Nepal earthquakes started in mid-March seen by satellite thermal observations (Figure 2). In April 20 atmospheric temperature was increased, which continued on Apr 21-22 with thermal filed build up on the top of the atmosphere (OLR) near the epicentral area (Figure 3). The ionosphere immediately reacts «Incorrect verb form: A singular subject of a relative clause always takes a singular verb.» to these changes in the electric properties of the ground layer measured by GPS/TEC over the epicenter areas, which have been confirmed as spatially localized increase in the DTEC on April 21 and 24 (Figure 3,4). The same scenario has occurred and for M7.3 of May 12, 2015.

Our preliminary results from recording atmospheric and ionospheric conditions during the M7.8 and M7.3 earthquake in Nepal using OLR monitoring on the top of the atmosphere, GIM-GPS/TEC maps, vertical TEC of lhaz GPS/GLONASS receiver in the region and atmospheric temperature from ground measurements show the presence of anomalies in the atmosphere, and ionosphere occurring consistently over region near the 2015 Nepal earthquake epicenters.

## Acknowledgments

The authors thank ISSI (Bern) for support of the team "Multi-instrument Space-Borne Observations and Validation of the Physical Model of the Lithosphere-Atmosphere-Ionosphere-Magnetosphere Coupling". We also thank NOAA/ National Weather Service National Centers for Environmental Prediction Climate Prediction Center for providing OLR and surface temperature data. The IONEX data and RINEX data (for lhaz GPS/GLONASS receiver in this study were acquired as part of NASA's Earth Science Data Systems and archived and distributed by the Crustal Dynamics Data Information System (CDDIS).

## References

Cervone, G.,Maekawa, S., Singh, R.P., Hayakawa, M., Kafatos, M., and Shvets, A., Surface latent heat flux and nighttime LF anomalies prior to the Mw=8.3 Tokachi-Oki earthquake, *Nat. Hazards Earth Syst. Sci.*, 6, 109-114, 2006

Davidenko D.V. Diagnostics of ionospheric disturbances over the seismo-hazardous regions (Ph.D. thesis), Fiodorov Institute of Applied Geophysics, Moscow, 2013, 147 p.

Gruber, A. and Krueger, A., The status of the NOAA outgoing longwave radiation dataset. *Bulletin of the American Meteorological Society* ,65, 958–962,1984

Hayakawa  M. (Editor), Earthquake Prediction Studies: Seismo Electromagnetics, TERRAPUB, Tokyo, Japa ,2013.

Kon, S., Nishihashi, M., Hattori, K., Ionospheric anomalies possibly associated with M>=6.0 earthquakes in the Japan area during 1998-2010: Case studies and statistical study, *Journal of Asian Earth Sciences*, doi:10.1016/j.jseases.2010.10.005 .

Liu, J. Y., Chuo, Y.J., Shan, S.J., Tsai, Y.B., Chen, Y.I., Pulinets S.A., and Yu S.B., Pre-earthquake ionospheric anomalies registered by continuous GPS TEC measurement, *Ann. Geophys.*, 22, 1585-1593,2004

Liu, D.: Anomalies analyses on satellite remote sensing OLR before Jiji earthquake of Taiwan Province, Geo-Information Science,2(1), 33–36, (in Chinese with English abstract),2000

Mehta, A., and J. Susskind, Outgoing Longwave Radiation from the TOVS Pathfinder Path A Data Set, *J.Geophys. Res.,* 104, NO. D10,12193-12212,1999.

Němec, F, Santolı́k, O., Parrot, M., and Berthelier, J.J. Spacecraft observations of electromagnetic perturbations connected with seismic activity*, Geophys. Res. Lett.*, 35, L05109, doi:10.1029/2007GL032517,2008

Ohring, G. and Gruber, A.: Satellite radiation observations and climatetheory, *Advance in Geophysics*., 25, 237–304, 1982.

Ouzounov D., Bryant, N., Logan, T., Pulinets, S., Taylor, P. Satellite thermal IR phenomena associated with some of the major earthquakes in 1999-2004, *Physics and Chemistry of the Earth*, 31,154-163, 2006

Ouzounov D., Liu, D., Kang,C. , Cervone, G., Kafatos, M., Taylor, P. Outgoing Long Wave Radiation Variability from IR Satellite Data Prior to Major Earthquakes, *Tectonophysics,* 431, 211-220,2007

Ouzounov D., S.Pulinets, M.Parrot, K.Tsybulya, P.Taylor, A. Baeza (2011a) The atmospheric response to M7.0 Haiti and M8.3 Chilean earthquakes revealed by joined analysis of satellite and ground data, *Geophysical Research Abstracts* Vol. 13, EGU2011-5195-1, 2011, EGU General Assembly

Ouzounov D. S.Pulinets, K.Hattori, M, Kafatos, P.Taylor (2011b) "Atmospheric Signals Associated with Major Earthquakes. A Multi-Sensor Approach, in the book "Frontier of Earthquake short-term prediction study", M Hayakawa, (Ed), Japan, 510-531

Ouzounov D., S. Pulinets, A. Romanov, A. Romanov Jr., K. Tsybulya, D.Davydenko, M. Kafatos and P. Taylor (2011c) Atmosphere-Ionosphere Response to the M9 Tohoku Earthquake Reviled by Joined Satellite and Ground Observations, Earthq. Sci, 24, 557–564

Ouzounov D. and S. Pulinets, Observation of pre- earthquake processes in atmosphere probably associated with the M7.8 in Nepal. Preliminary results. *International Workshop on Earthquake Preparation Process 2015 - Observation, Validation, Modeling, Forecasting - (IWEP2)*, May 29-30, 2015, Chiba, Japan

Parrot M. Anomalous seismic phenomena: View from space in Electromagnetic Phenomena Associated with Earthquakes (Ed. by M. Hayakawa), Transworld Research Network, 205-233, 2009

Pulinets S.A., Boyarchuk K.A., 2004. Ionospheric precursors of earthquakes. Springer, Berlin, Heidelberg, New York, 315 p.

Pulinets S., Ouzounov, D., Karelin, A., Boyarchuk, K., Pokhmelnykh, L. The Physical Nature of Thermal Anomalies Observed Before Strong Earthquakes, *Physics and Chemistry of the Earth*, 31, 143-153, 2006

Pulinets S., Kotsarenko, A, Ciraolo, L., Pulinets, I Special case of ionospheric day-to-day variability associated with earthquake preparation, *Adv. Space Res.*, 39, 970-977, 2007

Pulinets S.A., K.A. Boyarchuk, A.M. Lomonosov, V.V. Khegai, and J.Y. Liu, Ionospheric Precursors to Earthquakes: A Preliminary Analysis of the foF2 Critical Frequencies at Chung-Li Ground-Based Station for Vertical Sounding of the Ionosphere (Taiwan Island), Geomagnetism and Aeronomy, 2002, 42, No. 3, pp.508-513

Pulinets S.A., Bondur V.G., Tsidilina M.N., Gaponova M.V., Verification of the Concept of Seismoionospheric Relations under Quiet Heliogeomagnetic Conditions, Using the Wenchuan (China) Earthquake of May 12, 2008, as an Example, Geomagnetism and Aeronomy, 50, 231-242, 2010

Pulinets S.A., Ouzounov D.P., Davidenko D.V. Earthquake Prediction is Possible!? Integral technologies of multiparameter monitoring the geoeffective phenomena within the framework of integrated lithosphere-atmosphere-ionosphere coupling model, Moscow, *Trovant Publ.*, 2014, 144 p

Pulinets S., Ouzounov D., Karelin A. ,Davidenko D. (2015) Physical Bases of the Generation of Short_Term Earthquake Precursors: A Complex Model of Ionization Induced Geophysical Processes in the Lithosphere–Atmosphere–Ionosphere–Magnetosphere System, Geomagnetism and Aeronomy, 55, No. 4, pp. 521–538

Tramutoli, V., Cuomo V, Filizzola C., Pergola N., Pietrapertosa, C. Assessing the potential of thermal infrared satellite surveys for monitoring seismically active areas. The case of Kocaeli (İİzmit) earthquake, August 17th, 1999, *Remote Sensing of Environmen*t, 96, 409-426, 2005

Tramutoli V, C. Aliano, R. Corrado, C. Filizzola, N. Genzano, M. Lisi, G. Martinelli, N. Pergola. On the possible origin of Thermal Infrared Radiation (TIR) anomalies in earthquake-prone areas observed using Robust Satellite Techniques (RST). *Chemical Geology,* vol. 339*, 157-168*, 2013

Tronin, A., Hayakawa, M., and Molchanov, O. Thermal IR satellite data application for earthquake research in Japan and China*, J. Geodynamics*, 33, 519-534, 2002

Xiong P, X. H. Shen, Y. X. Bi, C. L. Kang, L. Z. Chen, F. Jing, and Y. Chen Study of outgoing longwave radiation anomalies associated with Haiti earthquake, *Nat. Hazards Earth Syst. Sci.*, 10, 2169–2178, 2010

Zolotov O.V. Earthquake effects in variations of the total electron content, PhD thesis, Murmansk, 2015, 146 p.

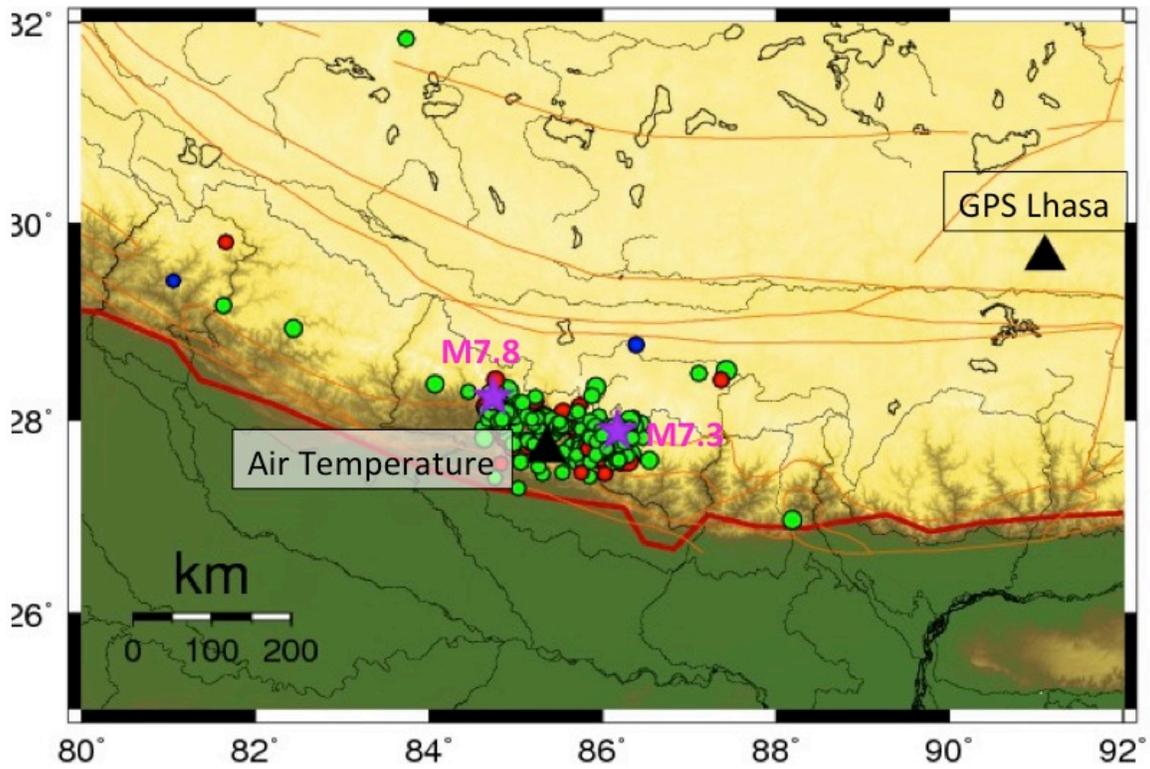

Figure 1. Reference map of Nepal region, with the location of earthquakes >M4 for Jan-May 2015. The location of M7.8 of April 25 and M7.3 of May12, 2015 are with purple stars. With black triangles are showing the location of the Air Temperature station (Katmandu) and GPS stations (Lhasa).

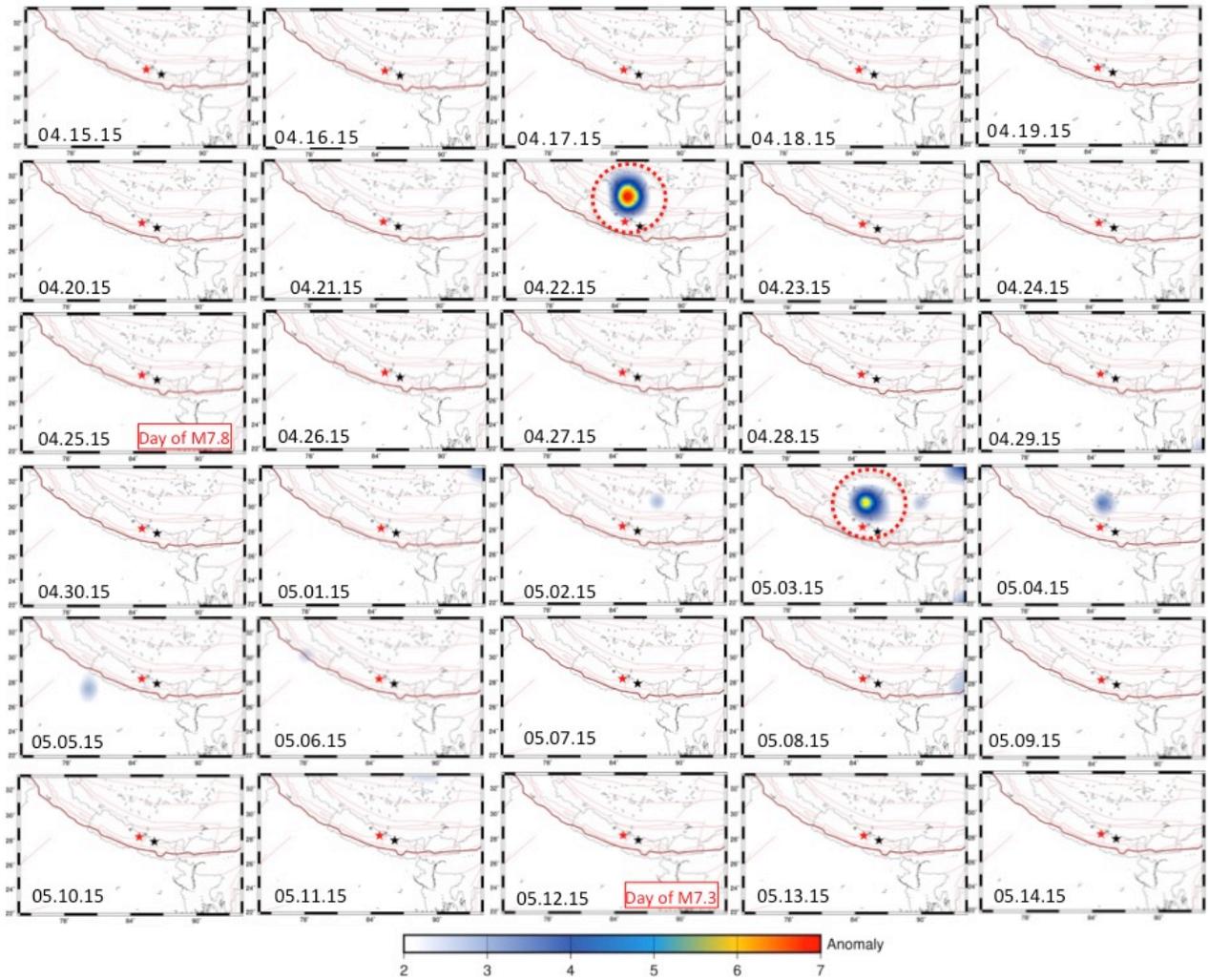

Figure 2. Time series of nighttime anomalous OLR observed from NOAA/AVHRR, April 15-May 14, 20115. Tectonic plate boundaries are indicated with red lines and major faults by brown ones and earthquake location by black stars. Red circle show the spatial location of abnormal OLR anomalies within vicinity of M7.8 (red star) and M7.3 (black star).

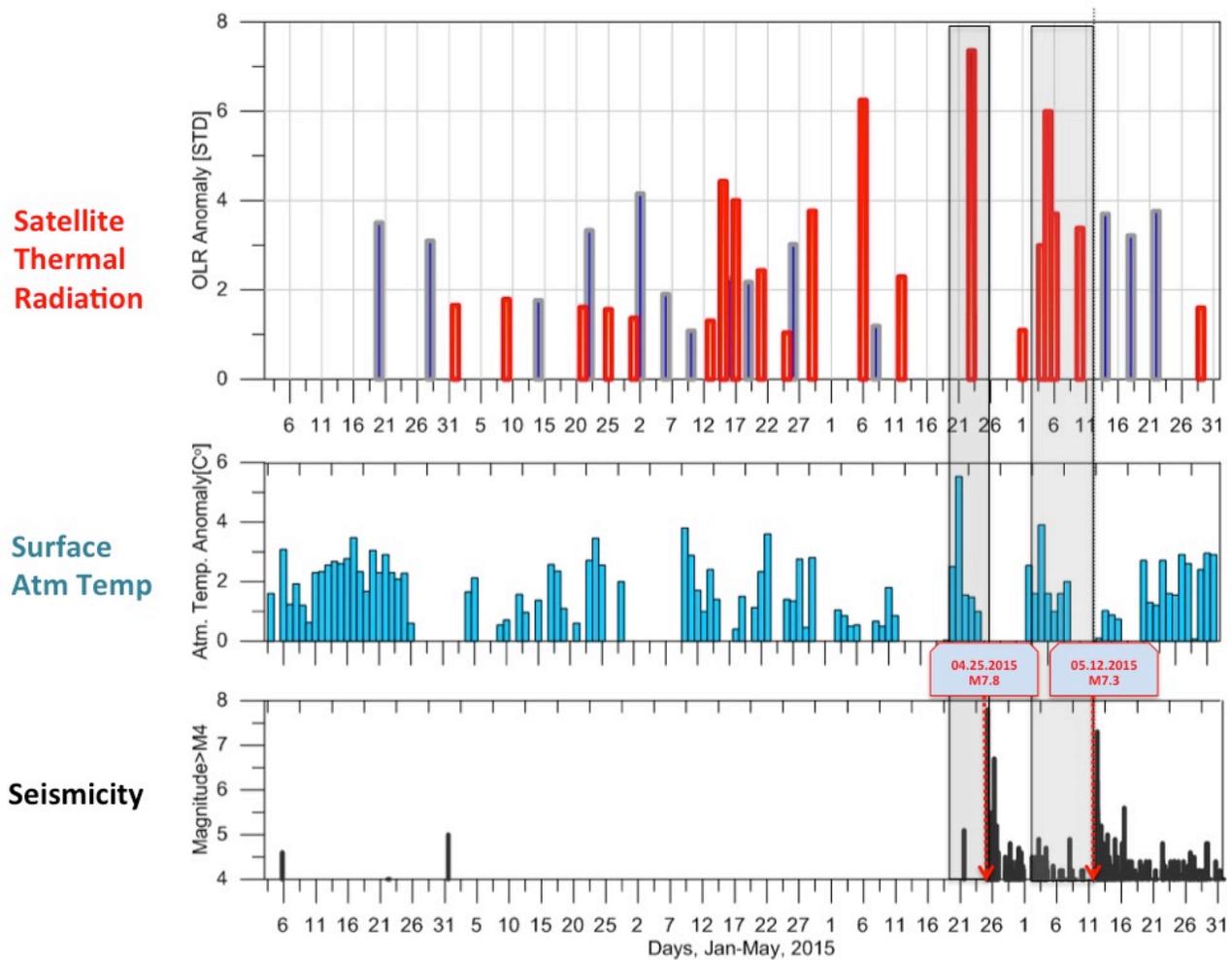

Figure 3. Time series of atmospheric variability observed within a 200 km radius of the Nepal earthquake (top to bottom): A) Nighttime anomalous OLR over epicentral region from January 1- May 31, 2015 observed from NOAA AVHRR (red). Same location, same period a year before – Jan-May 2014 (black); B) Air temperature anomaly from station Tribhuvan International Airport (blue) at 0600LT; C) Seismicity (M>4.0), Jan-May 2015 within 200km radius of the M 7.8 epicenter

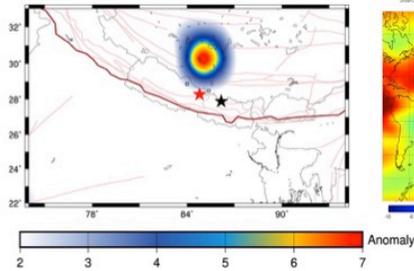 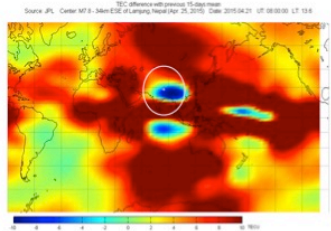 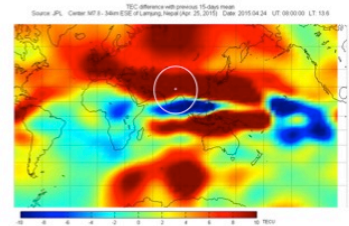

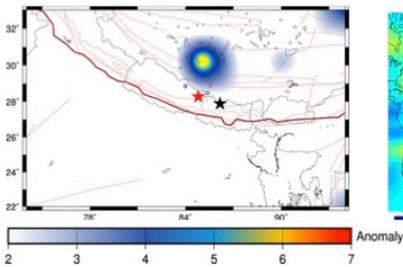 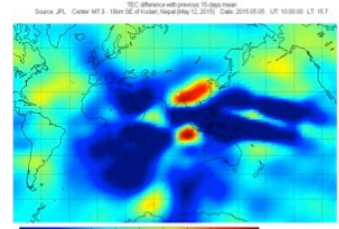 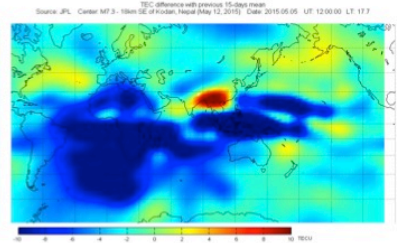

Figure 4. Satellite Thermal and GIM GPS/TEC spatial analysis.

Top panel: M7.8 of April 25, 2015,Nepal; (A) Satellite Thermal nighttime map of April 21-22 (mean), 06UT (-3 days), 2015; (B) Differential TEC map of April 21, 2015, (-4 days) 09UT and April 24, 2015, 08UT (-1 day)

Bottom Panel: M7.3 of May 12, 2015, Nepal; A) Satellite Thermal nighttime map of May 3-4 (mean), 06UT 2015 (-8 days); (B) Differential TEC map of May 5, 2015, (-7 days) 10UT and May 5, 2015, 12UT (-3 day)

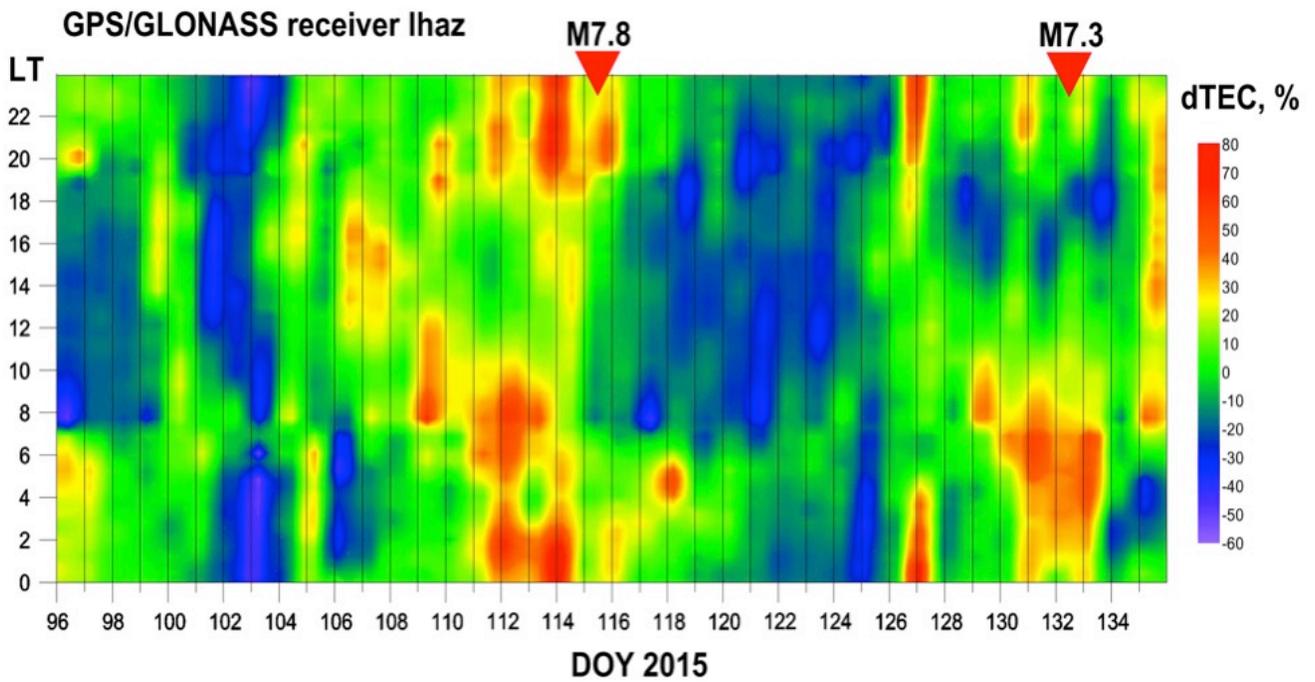

Figure 5. DTEC variations for the lhaz GPS/GLONASS receiver (Fig.1) for April 6 - May 15, 2015.